\documentclass[aps,superscriptaddress,nofootinbib]{revtex4}
\usepackage{graphicx}
\usepackage{amssymb,amsfonts,amsmath}
\usepackage{dcolumn}
\usepackage{natbib}
\usepackage[all]{xy}
\usepackage{booktabs}
\usepackage{multirow}
\usepackage{amssymb}
\usepackage{booktabs}
\usepackage{multirow}
\usepackage{color}
\usepackage{amsmath,epsfig}

\newcommand{\rem}[1]{{\color{magenta} #1}}

\begin{document}

%\title{Statistically validated mobile communication networks: A comparison of European and Chinese data}

\title{Statistically validated mobile communication networks: Evolution of motifs in European and Chinese data}

\author{Ming-Xia Li}
\affiliation{School of Business, School of Science and Research Center for Econophysics, East China University of Science and Technology, Shanghai 200237, China}

\author{Vasyl Palchykov}
\affiliation{Dept. of Biomedical Engineering and Computational Science, Aalto University, FI-00076 Aalto, Finland}
\affiliation{Institute for Condensed Matter Physics, National Academy of Sciences of Ukraine, UA-79011 Lviv, Ukraine}
\affiliation{Instituut-Lorentz, Universiteit Leiden, 2300 RA Leiden, The Netherlands}

\author{Zhi-Qiang Jiang}
\affiliation{School of Business, School of Science and Research Center for Econophysics, East China University of Science and Technology, Shanghai 200237, China}

\author{Kimmo Kaski}
\affiliation{Dept. of Biomedical Engineering and Computational Science, Aalto University, FI-00076 Aalto, Finland}

\author{Janos Kertesz}
\affiliation{Center for Network Science, Central European University, Nador 9, H-1051, Budapest, Hungary}
\affiliation{Dept. of Biomedical Engineering and Computational Science, Aalto University, FI-00076 Aalto, Finland}

\author{Salvatore Micciche}
\affiliation{Dipartimento di Fisica e Chimica, Universit\`a di Palermo, Viale delle Scienze, Ed. 18, I-90128, Palermo, Italy}

\author{Michele Tumminello}
\affiliation{Dipartimento di Scienze Economiche, Aziendali e Statistiche, Universit\`a degli Studi di Palermo, Viale delle Scienze, Edificio 13, I-90128 Palermo, Italy}

\author{Wei-Xing Zhou}
\affiliation{School of Business, School of Science and Research Center for Econophysics, East China University of Science and Technology, Shanghai 200237, China}

\author{Rosario N. Mantegna}
\affiliation{Center for Network Science, Central European University, Nador 9, H-1051, Budapest, Hungary}
\affiliation{Dipartimento di Fisica e Chimica, Universit\`a di Palermo, Viale delle Scienze, Ed. 18, I-90128, Palermo, Italy}
\affiliation{Department of Economics, Central European University, Nador 9, H-1051, Budapest, Hungary}

\date{\today}

\begin{abstract}

Big data open up unprecedented opportunities to investigate complex systems including the society. In particular, communication data serve as major sources for computational social sciences but they have to be cleaned and filtered as they may contain spurious information due to recording errors as well as interactions, like commercial and marketing activities, not directly related to the social network. 
%However, communication data, one of the major sources for computational social sciences have to be cleaned and filtered as they may contain spurious information due to recording errors and interactions not related to the social network, like commercial and marketing activities.
The network constructed from communication data can only be considered as a proxy for the network of social relationships. Here we apply a systematic method, based on multiple hypothesis testing, to statistically validate the links and then construct the corresponding Bonferroni network, generalized to the directed case. We study two large datasets %The study is done for two large datasets 
of mobile phone records, one from Europe and the other from China. For both datasets we compare %We compare 
the raw data networks with the corresponding Bonferroni networks and point out significant differences in the structures and in the basic network measures.     %for both datasets. 
We show evidence that the Bonferroni network provides a better proxy for the network of social interactions than the original one. By using the filtered networks we investigated the statistics and temporal evolution of small directed 3-motifs and conclude that closed communication triads have a formation time-scale, which is quite fast and typically intraday. We also find that open communication triads preferentially evolve to other open triads with a higher fraction of reciprocated calls. These stylized facts were observed for both datasets.
%We use the filtered networks to investigate the statistics and temporal evolution of small directed 3-motifs. We conclude that closed communication triads have a formation timescale which is quite fast and typically intraday. Open communication triads preferentially evolve to other open triads with a higher fraction of reciprocated calls. These stylized facts are observed in both datasets. 
\end{abstract}
\maketitle

%%%%%%%%%%%%%%%%%%%%%%%%%%%%%%%%%%%%%%%%%%%%%%%%%%%%%%%%%%%%%%%%

\section{Introduction}
The data deluge has its origin in the development of information communication technology, which in turn has revolutionized the scientific research of social systems. The ``digital footprints'' we leave behind in almost all of our activities enable unprecedented investigations both in depth and sample sizes. A new discipline, called ``computational social science'' \cite{Lazer2009}, has emerged to join the efforts of social scientists, computer scientists, physicists, and mathematicians in a truly interdisciplinary approach with the aim to better understand the laws of human society both at an individual and at a collective level. %of understanding better the laws of human society both at an individual and at a collective level. 

Mobile call records (MCRs) play a special role in the studies of human societies as the mobile phone coverage is close to 100 \% in the adult population and these equipments are our companions in almost all of our activities. Accordingly, MCRs are well suited to map out the structure of social networks \cite{Onnela2007, Eagle2009} including the dynamics \cite{Palla2007} and hierarchical structure \cite{Blondel2008} of the communities, to study dynamic aspects of human behavior like mobility characteristics \cite{Song2010, Lu2012, Schneider2013} or communication patterns \cite{Karsai2011, Kivela2012} and temporal motifs \cite{Kovanen2011}. The origin of samples shows an increasing  socio-economic and cultural variety ranging from different European sources \cite{Onnela2007, Blondel2008} to American \cite{Lu2012}, Asian \cite{Jianga2013} and African \cite{Blondel2012} data sets that have been investigated. Although no systematic comparison have yet been made, some universal features seem to emerge. These include the Granovetterian structure \cite{Onnela2007,Granovetter1973} of the network, the intermittent, bursty character of communication \cite{Barabasi2005}, and the strong inhomogeneity both in the number of contacts and in the strength of the activities \cite{Onnela2007b}.

MCRs indeed provide detailed information about human interactions: data for millions of users about who called whom, when, how long the conversation took and the whereabouts of the callers serve as a gold mine to understand the structure of the society and the dynamic laws of communication and mobility of individuals. Moreover, if so-called metadata are 
at the disposal of the researchers, deeper insight about gender and age related behavioral patterns can be mapped out \cite{Palchykov2012, Kovanen2013}.

Mesoscopic structures like network motifs \cite{Alon2002} are of particular interest for the understanding of the structure and function of the society. A motif is a set of isomorphic subgraphs and it is generally assumed that it represents a functionally important group of nodes if its cardinality significantly exceeds the expected number of such subgraphs in a reference system, being usually the configuration model. %which is usually the configuration model. Using 
With this concept, it was possible to identify classes of networks, where within one class (e.g., networks of genetic transcription in different species or networks of different languages) there is %have 
similar overly representation of motifs. Based on weight intensity the concept of motifs was also generalized to weighted networks \cite{Onnela2005}. Recently, there has been growing interest in the dynamic patterns. Dynamic motifs \cite{Kovanen2011,Kovanen2013} are classes of similar event sequences, where the similarity refers not only to topology but also to the temporal order of the events (e.g., phone calls). Using the information about the locality of the calls mobility motifs were defined and the classification of human mobility patterns was enabled \cite{Gonzales2013}.

In temporal networks \cite{Holme2013}, where links are present only temporarily for an interaction event, static motifs defined on the aggregate networks present a time evolution as a function of the aggregation window. So far this has not been investigated, although this dynamics contains interesting information about the system. Time stamped mobile phone data are particularly suitable for such a study. One then asks what are the typical motifs that are over-expressed in the network of MCRs? How do they emerge as a function of time? What is the characteristic time needed for the evolution of the motifs? These are the questions we will focus on in this paper. 

In social science there is a long tradition for studying triads, i.e. subgraphs of 3 nodes connected by directed links \cite{Wasserman}. Recently, triads have been investigated in many other classes of complex networks ranging from biological to economic and financial networks \cite{Squartini2013,Bargigli2013}. 
Following the terminology introduced by U. Alon and collaborators, triads are called 3-motifs, and motifs of higher order (typically of order 4 or 5) are also considered. The most developed theories and the largest number of empirical studies about motifs concern 3-motifs. In fact the number of different motifs explodes when motifs of higher order are considered and the investigation of 4- or 5-motifs in large networks is extremely demanding from computational point of view. In social sciences a large amount of studies have focused on the 3-motif statistics and dynamics with the aim to use this information to detect global properties of the investigated social system. So far the most prominent property investigated in social networks is triadic closure. The triadic closure is observed when a triad with only two relationships detected among the three social actors evolves to another triad with all the pairwise relationships present to some degree \cite{Wasserman}.

In the present study we investigate the 3-motifs observed in two large databases of MCRs. Specifically we investigate MCRs of two different mobile companies one operating in Europe and another one 
in China. In this way we are studying the effectiveness of our approach and the validity of our investigation of communication links of social origin in two datasets that are different in various respects such as the telecom company (with its specific commercial policy), the geographical location and the recording time period. We have chosen to focus our investigation on the directed 3-motifs for two main reasons: (i) because they are extremely informative from a social point of view and (ii) because a reliable empirical estimation requires a series of strict conditions in the processing of very large samples. 

A major problem with big data is that they have to be cleaned and filtered as they contain spurious information due to recording errors and interactions, like commercial and marketing activities, not directly related to the study at hand.
Usually MCRs are not collected for scientific purposes and even if the companies attempt to provide the relevant data there could be serious problems. One example is that for studying social relationships private communication is needed, however, the experience tells that sometimes phones registered as private are used for professional purposes like call centers or marketing and information campaigns. In fact, the presence of large spurious communication hubs, e.g. large call centers, significantly alters the statistics of 3-motifs (and, more generally, of any class of motifs). Dialling wrong numbers is another possible source of false links. In addition, usual corruption during coding, transferring and processing data can take place. Unless data are cleaned, spurious links could be misinterpreted as real social relationships. This problem is part of the general topic of information filtering in complex networks, with strong inhomogeneities \cite{Radicchi2011}. In fact, human related systems usually show properties changing over many orders of magnitude and this is so for communication networks also: The distributions of degrees or activities are fat tailed \cite{Onnela2007b}.

A somewhat arbitrary way of filtering data was introduced by Onnela et al. \cite{Onnela2007}. Three measures were taken: i) only mutual connections were considered as links, i.e., both individuals had to initialize call during the period of observation; ii) links with total call duration of less than 10 seconds during examined period of 18 weeks were ignored; and iii) the nodes with less than 60 seconds of total call activity were filtered out. In fact, this way spurious nodes with enormous ($~10^4$) number of unidirectional connections and sometimes more than 24 hours/day (!) activity were eliminated. The 10s cutoff served to filter out the calls of wrong numbers. However, this method unintentionally %unwillingly
distorts the results as there can be many socially relevant unidirectional links and even short duration links that may carry social interaction.

Another, more systematic way of filtering was proposed by Serrano et al. \cite{Serrano2009}. The idea is to statistically validate the links by deciding locally, which of the links carry disproportionate fraction of the weights adjacent to a given node. Comparing the empirical observations with a null model that takes into account the inhomogeneities of the system, can show significant over-representation of links
%deviations from the null model 
thus indicating %indicate 
their relevance. Carrying out the procedure node-by-node results in what is called the ``multiscale backbone'' of the system. This method points out important aspects of filtering including the necessity of statistical validation and the relevance of the appropriate null model. However, it is asymmetric for the nodes of the links and it has some restrictions upon %, it has some restrictions upon
the degree $k$ of the nodes (isolated links between two $k=1$ nodes can never be validated, irrespective of the weight of the link) and it handles the local network topology independently of the rest of the network.

Recently, a method to filter out statistically significant links in bipartite complex networks \cite{Tumminello2011} was proposed. As the mobile call network can be considered as a bipartite one, where one set of nodes corresponds to the mobile phone users and the other one to the calls they perform, the method can be straightforwardly applied to our problem. As it is based on multiple hypothesis testing, global information is built in, thus the above-mentioned problems can be avoided. This method has already been applied successfully to a number of systems, including the networks of organisms, financial stocks and the Internet Movie Database \cite{Tumminello2011}, classification of investor strategies \cite{Tumminello2012} and of specialization of criminal suspects \cite{Tumminello2013}. 

In this paper we adapt and apply the method introduced in ref. \cite{Tumminello2011} to mobile phone communication networks from two different regions, which are a European country and the province level municipality of Shanghai (China). We first construct the communication networks from the raw MCRs (``original networks'') and then the networks of the statistically validated links, also called Bonferroni networks. We keep the directed character of the links as they carry important information about the relationships. The comparison between the original and the Bonferroni networks shows significant differences in the basic statistical properties. For example, our filtering removes the extremely  large hubs (which would contradict the social brain hypothesis \cite{Dunbar1998}) but keep a large number of unidirectional contacts. 

The Bonferroni filtering of the original network allows us to perform a detailed analysis of so-called 3-motifs. We show that the study of 3-motif statistics and dynamics is unreliable unless we perform the Bonferroni filtering to the original network. This is due to the fact that the empirical estimation of 3-motifs is strongly affected by the presence of huge communication hubs that have no social origin but only some socio-technical motivation such as is the case of call centers. In the Bonferroni network, we study the time evolution of the communication 3-motifs. Our results show that communication 3-motifs are typically characterized by triadic closure at an intraday time scale. In fact, 3-motifs with only links detected between two pairs of subscribers primarily evolve into other 3-motifs characterized by a higher number of reciprocated calls. Triadic closure is preferentially observed in communication 3-motifs only after the calls of the open 3-motif are reciprocated.

The paper is structured as follows. In the next section we discuss the application of the Bonferroni network method to the MCRs. In Section III we describe our results for the directed 3-motifs. Then the temporal evolution of the motifs is discussed. Finally we present the conclusions.

\color{black}{
\section {Bonferroni network of mobile call records}}

In order to analyze communication data for constructing MCR based networks one has to first decide whether the entries in the records serve as good proxies for real social interactions in a probabilistic sense. This is a multiple hypothesis test validation problem, which we  approach 
by adapting and applying a directional version \cite{Hatzopoulos2013} of the recently introduced method of Bonferroni networks \cite{Tumminello2011}.\\

\subsection{Data}
We investigate two sets of data: One from a Chinese mobile phone service provider and another one from a European service provider. The Chinese data contain time stamped data of all (hashed) subscribers of the service within the time periods from June 28 to July 24 2010, and from October 1 to December 31 2010. In the second period, the calls recorded on October 12, November 5, 6, 13, 21, and 27, and December 6, 8, 21, and 22 contain missing records and these days are removed from %in 
our analysis. Thus we have in total %totally have 
109 days of calls recorded for Chinese data. This data set consists of 4,031,090 subscribers and 1,091,695,590 calls (done with both subscribers and non-subscribers of the service provider). When we select calls occurring only among subscribers, the number of calls reduces to 128,410,897, i.e., 88.24\% of the calls go to non-subscribers. The set of mobile phone users including subscribers and non-subscribers exceed nine millions. 

The data from the European provider contain all records of its 7,387,034 subscribers during 212 days between January 1, 2007 and July 31, 2007. This includes 3,969,043,426 calls, 682,124,009  of which occurred  between subscribers of the given provider, i.e., %occurred between subscribers of given provider, i.e. 
81.54\% of all calls connected subscribers with non-subscribers. The whole set of subscribers and non-subscribers exceeds 91 million users.

By considering that the primary focus of our investigations is on the evolution of 3-motifs, in the present study we perform our investigations on the calling networks of subscribers only. In fact, including non-subscribers would alter the 3-motif statistics because calling data between two non-subscribers are not recorded in our datasets.

\subsection{Statistically Validated Networks}

In our directed network nodes are mobile phone subscribers and a directional link is set from subscriber A to subscriber B if A makes a call to B in a selected time window, the weight $w$ of the link being the number of calls in the investigated time window. For each link in the network, we perform a statistical test to check whether a link is statistically validated against a null hypothesis assuming heterogeneous calling profiles of the subscribers. The method is a directional variant of the method introduced in ref. \cite{Tumminello2011}. The statistical test is implemented as follows. We define $N$ as the total number of calls among the subscribers in the system, and focus on two subscribers, $i$ and $j$, to check whether the number of calls of $i$ to $j$ are over-expressed 
with respect to a null hypothesis taking into account heterogeneity in the number of performed calls. Let us call $n_{c}^i$ the number of times in which subscriber $i$ makes a call, and $n_{r}^j$ the number of times in which subscriber $j$ receives  a call. By labelling with %as 
the number of times subscriber $i$ calls $j$ with $X$, the probability of observing X calls is given by

$$
H(X|N,n_{c}^i,n_{r}^j)=\frac{{n_c^i \choose X} \, {N-n_c^i \choose n_r^j-X}}{{N \choose n_r^j}}.
$$

We can therefore associate a $p$-value with the observed number $X=n_{cr}^{ij}$ of calls from subscriber $i$ to subscriber $j$ as

$$p(n_{cr}^{ij}) =\sum_{X=n_{cr}^{ij}}^{{\rm min}[n_{c}^i,n_{r}^j]} H(X|N,n_{c}^i,n_{r}^j).$$

Calculating the $p$-value for all the directed edges, $N_E$ in our network, implies that we run $N_E$ statistical tests for obtaining the network. When a large number of statistical tests are performed simultaneously the effectiveness of the statistical test can be decreased by a large number of false positive unless a multiple hypothesis test correction is used.  In the present study we use the Bonferroni correction, which is the strictest multiple hypothesis test correction controlling the familywise error rate when either dependent or independent multiple hypotheses are tested. That means that the univariate level of statistical significance $p_u=0.01$ must be corrected, and the multivariate level is $p_m = p_u/N_E=0.01/N_E$. 

If the estimated $p(n_{cr}^{ij})$ is less than $p_m$, we conclude that the link from subscriber $i$ calling subscriber $j$ is not due to the high heterogeneity of the subscribers and most probably reflects a social interaction between the subscribers. Accordingly, we set a link from $i$ to $j$ in the filtered network that is named Bonferroni Network.

\begin{figure}[ht]
\begin{center}
\includegraphics[scale=0.4]{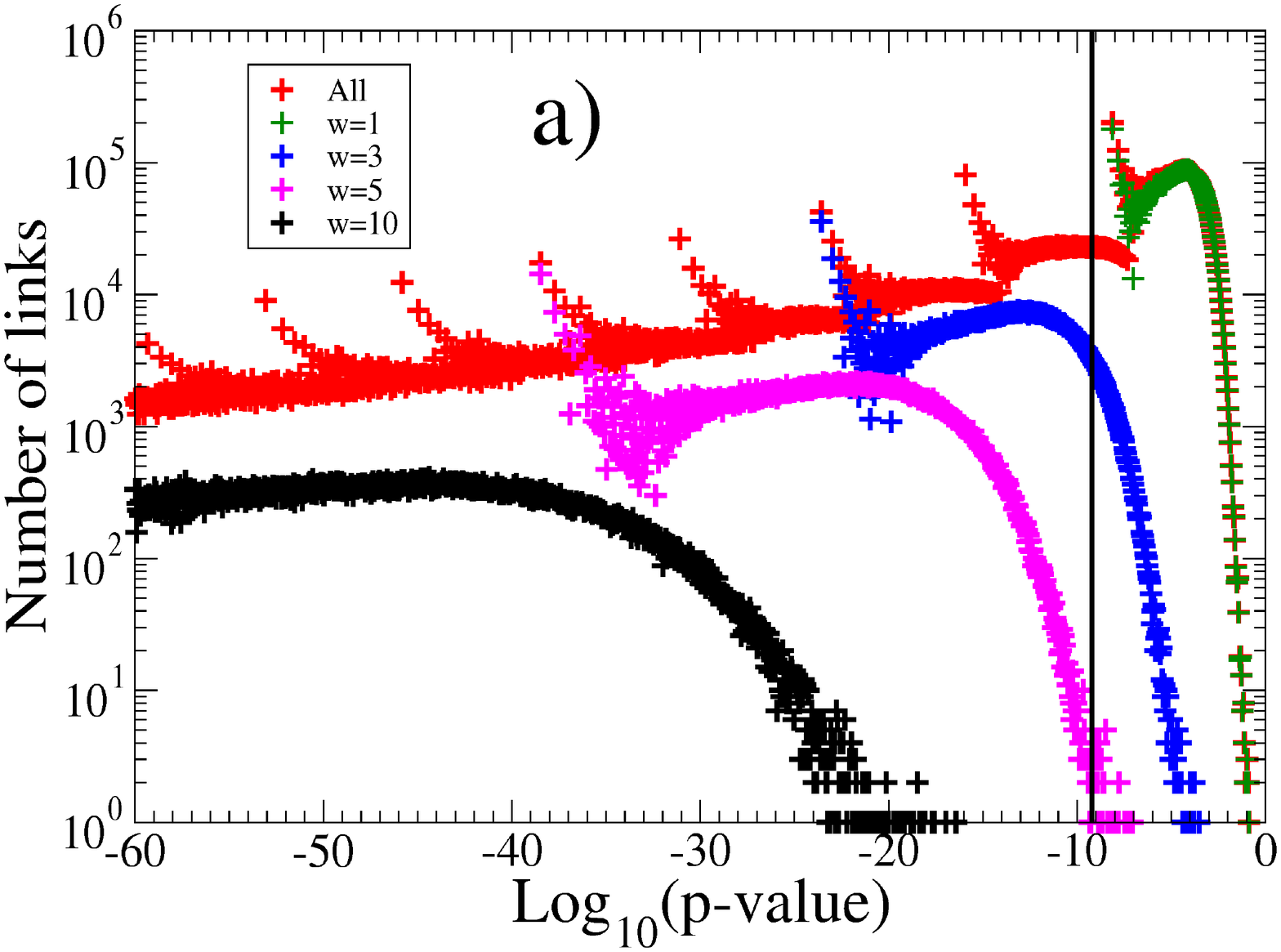} 
\includegraphics[scale=0.4]{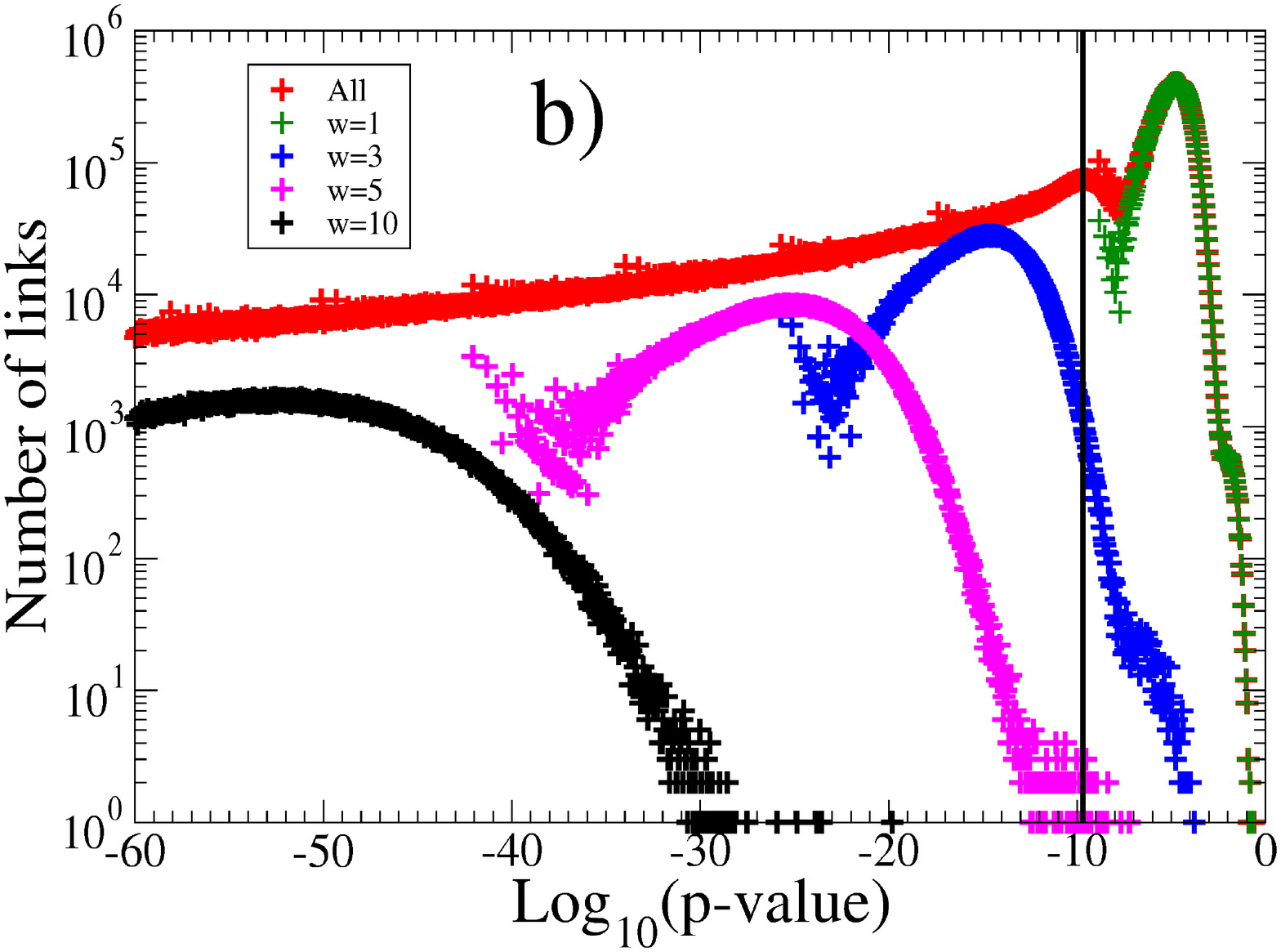}\\
\caption{\label{fig1}
Number of links as a function of the $p$-value for the Chinese (panel a) and European (panel b) datasets. The red symbols describe the histogram for all links. Symbols of different color refer to the number of links of pairs of callers and receivers with weight equals to 1 (green), 3 (blue), 5 (purple) and 10 (black). The vertical line indicates the Bonferroni threshold. Links located to the left of the threshold are retained in the Bonferroni network. The network is obtained by considering  the entire period. Only links between subscribers are considered. }  
\end{center}
\end{figure}

In Fig. \ref{fig1} we show a series of histograms of the number of links characterized by a certain $p$-value for the Chinese and the European datasets respectively. Different histograms (characterized by different colors) are obtained by grouping the various links in terms of the number of calls characterizing them, i.e., in terms of their weights. The time interval used to build the original network is the entire time period available that is 109 days and 212 days for the Chinese and European dataset respectively. Figure\ref{fig1}     
shows that in both datasets the links with just one call (weight equals to one) are characterized by a $p$-value which is larger than the Bonferroni threshold (indicated as a vertical line). The links that are filtered from the original network comprise essentially all the links with weight one and some of the links with weight up to five. Under the conditions of our analysis, both for the Chinese and the European datasets, when the weight is larger than five the links are always included in the Bonferroni network (see the case of $w=10$ in Fig. \ref{fig1}).

The fact that links with unit weight are not present in the Bonferroni network is due to the statistical validation procedure of the links. In fact, for unitary weights the above defined $p$-value reads:
\begin{eqnarray}
                           p_1 =\sum_{X=1}^{\rem {m}in[n_{c}^i,n_{r}^j]} H(X|N,n_{c}^i,n_{r}^j) = 1-H(0|N,n_{c}^i,n_{r}^j) \ge& 1-H(0|N,1,1)=H(1|N,1,1)=\\
= & \frac{1}{N} \ge \frac{0.01}{N_E} = p_m \Leftrightarrow 100 \ge \frac{N}{N_E},
\end{eqnarray}
and the last inequality holds true in our system, because the average number of calls per directed link, $\frac{N}{N_E}$, is much smaller than 100.
However, as mentioned above, with our statistical validation procedure also some of the links with higher weights do not get validated in the Bonferroni network. The absence of validation is not a direct consequence of the small average number of phone calls per link. For example,  let us consider a simple case in which $n_{r}^j=n_{c}^i=n_{cr}^{ij}=2$. In this case, the $p$-value is
\begin{equation}
p_2=\sum_{X=2}^{2} H(X|N,2,2)=H(2|N,2,2)=\frac{2}{N\,(N-1)}. 
\end{equation}
This $p$-value would be statistically significant if it were $< 0.01/N_E$, and such a condition is easily attained, even in a sparse system like the present one,
 \begin{equation}
p_2=\frac{2}{N\,(N-1)} < \frac{0.01}{N_E} \Leftrightarrow \frac{200}{N-1} < \frac{N}{N_E}.
\end{equation}
Indeed, the latter inequality says that, to validate the link from $i$ to $j$ is just sufficient that the average number of calls per link is larger than $\frac{200}{N-1}$, which is a quantity smaller than 1 in any setting that includes more than 201 phone calls.

In the setting of the Bonferroni threshold there is a margin of arbitrariness. In other words, which is the most appropriate threshold to be used when we obtain distinct daily networks that we wish to compare? We believe that the answer to this question depends on the type of comparisons that one aims to perform on the obtained networks. Therefore there might be more and less restrictive choices in the setting of the Bonferroni threshold. To maintain the potential number of false positive link selections, in the present study we set the Bonferroni threshold to $p_m =0.01/N_E$, where $N_E$ is the number of edges observed in the investigated periods (we have a single investigated period for the European data and 2 distinct investigated periods in the Chinese data \footnote{For the Chinese dataset, we have two periods of data. In the first period (first month) 6,441,490 links are present in the original network between 2,309,619 subscribers  and in the second period (last three month) 13,616,634 links are present between 3,492,116 subscribers . We consider the two time periods as separate time periods and we set  two Bonferroni thresholds as 0.01/6,441,490 and 0.01/13,616,634 when performing the construction of the Bonferroni networks.  For the original daily networks, we have about $6.53 \cdot 10^5$ nodes and $6.94 \cdot 10^5$  links on average. By using the statistical test, 52.17\% nodes and 65.87\% links are removed in daily networks on average.}). In this way the Bonferroni threshold is rather conservative for the networks computed at short time intervals, e.g., at daily and weekly time intervals.

It should be noted that the choice of the Bonferroni threshold within broad limits up to some orders of magnitude does not crucially affect the composition of the networks obtained at the different time intervals. For instance, if the Bonferroni threshold used to construct daily Bonferroni networks for the Chinese data is increased by one order of magnitude, then on average the number of statistically validated links increases by $4.8. \%$.

\subsection{Basic metrics and degree distribution}

\begin{figure}
\begin{center}
\includegraphics[scale=0.9]{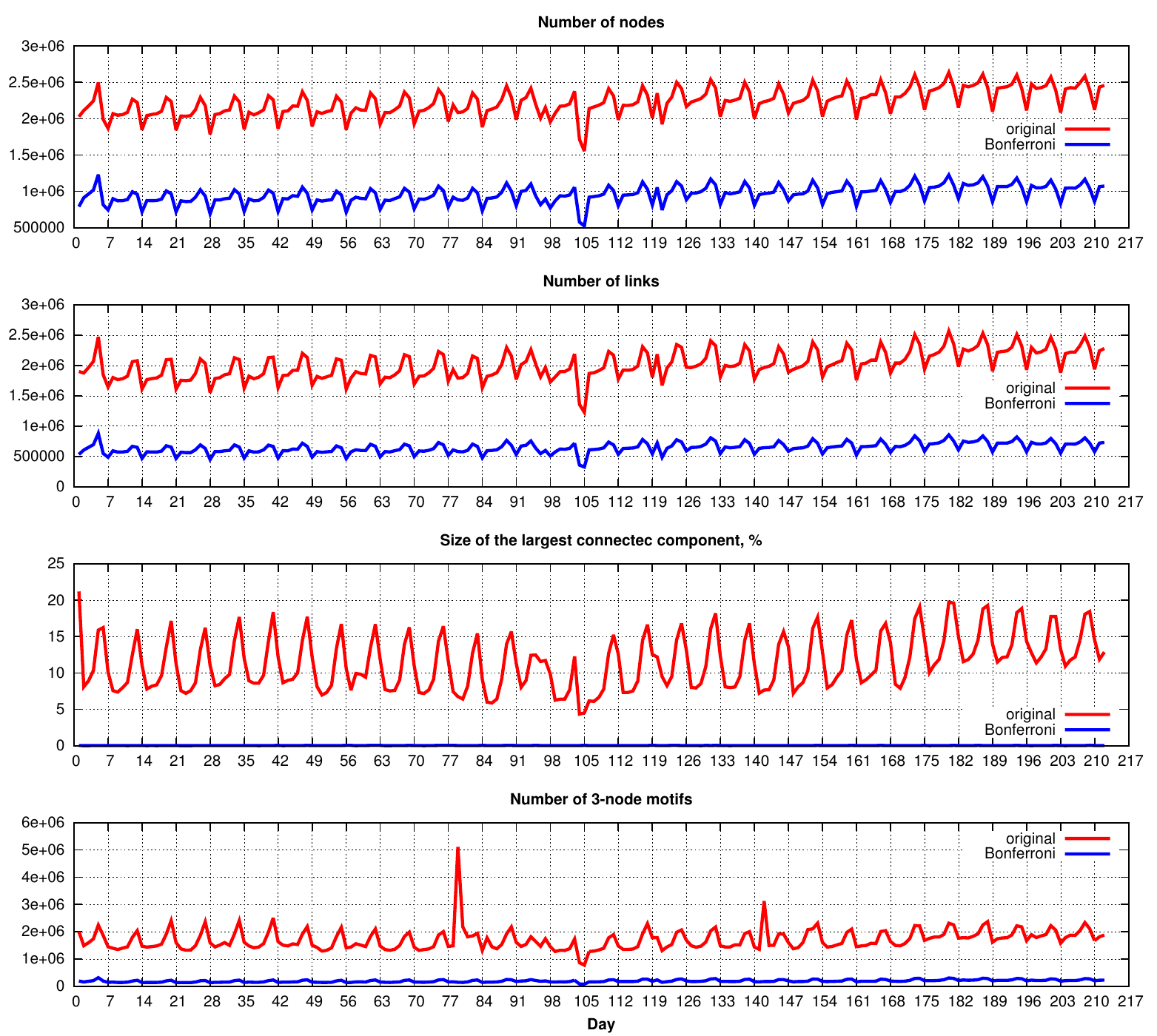}
\caption{Time evolution of some basic network indicators for the daily networks obtained from MCRs of the European dataset. The top panel shows the number of nodes (subscribers) which are present in the original (red line) and in the Bonferroni (blue line) network. A weekly pattern is clearly seen in both curves and some special days are also observed (see for example day 5 and 105). The middle top panel shows the number of links. The daily Bonferroni networks retain roughly one third of the links. The middle bottom panel shows the percent of nodes present in the largest connected component. The original network has a largest connected component including on average 12\% of the nodes. A large weekly cycle is observed. The largest connected component of the Bonferroni network has a negligible fraction of the nodes. The bottom panel shows the time evolution of the number of 3-motifs. In the original network this number is fluctuating and present a huge spike at day 79. In the case of the Bonferroni network, the time evolution is fluctuating less and the huge spike is not present.} 
\label{fig3} 
\end{center}
\end{figure}

In Fig. \ref{fig3} we show the time evolution of some basic network indicators for the daily networks obtained from MCRs of the European dataset. The top panel shows the number of nodes (subscribers) which are present in the original (red line) and in the Bonferroni (blue line) network. Roughly half of the nodes present in the original network are also present in the Bonferroni network. A weekly pattern is clearly seen in both curves and some special days are also observed (see for example day 5 and 105). The middle top panel shows the number of links. For the investigated set and for the considered time period, the Bonferroni network retains roughly one third of the links of the original network. The middle bottom panel shows the percent of nodes present in the largest connected component. On average the original network has a largest connected component including 12\% of the nodes. A large weekly cycle of amplitude close to 5\% is observed. It is worth noting that the largest connected component of the Bonferroni network has a negligible fraction of the nodes. This means that the Bonferroni network shows a large number of disconnected clusters of subscribers without a giant component. The interconnection is provided by the presence of large hubs and weak ties that are filtered out in our approach.

This behavior is specific to the daily networks.  The percent of nodes in the largest connected component increases, when the period of time used to detect the network increases. In Table \ref{TableA} we show the average percent and the standard deviation of the largest connected component of original and Bonferroni networks obtained for different time periods for the Chinese and European datasets. From the table we see that already for weekly networks the percent of nodes of the largest connected component is almost 49\% and 35\% for the Chinese and European datasets respectively. These values further increase for the monthly networks when the largest connected components of the Bonferroni networks are 72\% and 81\%, i.e., values not too different from the ones (81\% and 94\%) observed in the original networks for the Chinese and European datasets respectively. We interpret these results as an indication that our filtering methodology is able to detect a progressively increasing fraction of the weak ties that provide the interconnections building the largest connected component. This observation is in agreement with the detected role of weak ties discussed in Ref. \cite{Onnela2007}.

\begin{table}[ht]                                                                                                         
\caption{Summary of the average value of the percent of nodes present in the largest connected component of daily, weekly, monthly and complete networks both for the networks of the original set and for Bonferroni networks. We also report the standard deviation (in percent) of the average value. The results were obtained for the Chinese (top) and European (bottom) datasets. Subscribers only.}                                                       
\centering                                                                                                                
\begin{tabular}{*{11}{c}}                                                                                                 
\hline\hline                                                                                                              
 & & \multicolumn{4}{c}{Original}                                                                                         
 & & \multicolumn{4}{c}{Bonferroni}\\                                                                                     
\cline{3-6}
\cline{8-11}                                                                                                   
 & & Daily & Weekly & Monthly & Complete set                                                            
 & & Daily & Weekly & Monthly & Complete set                                                            
% Chinese data
\\                                                                                                                        
\hline                                                                                                                    
 & Mean (\%)                                                                                                                 
 & 29.33                                                                                                                  
 & 68.62                                                                                                                  
 & 81.47                                                                                                                  
 & 85.69                                                                                                                 
 &                                                                                                                        
 & 0.41                                                                                                                  
 & 48.96                                                                                                                   
 & 72.50                                                                                                                 
 & 79.12                                                                                                                 
\\                                                                                                                        
& SD (\%)                                                                                                                 
 & 8.53                                                                                                                  
 & 2.66                                                                                                                  
 & 2.55                                                                                                                  
 & -                                                                                                            
 &                                                                                                                        
 & 0.26                                                                                                                  
 & 4.82                                                                                                            
 & 2.47                                                                                                                 
 & -                                                                                                                 
\\                                                                                                                        
\hline
% European data                                                                                                                        
\hline                                                                                                                    
 & Mean (\%)                                                                                                                 
 & 11.45                                                                                                                  
 & 75.85                                                                                                                  
 & 93.89                                                                                                                  
 & 98.85                                                                                                                 
 &                                                                                                                        
 & 0.018                                                                                                                  
 & 34.53                                                                                                                   
 & 81.35                                                                                                                 
 & 96.79                                                                                                                 
\\                                                                                                                        
& SD (\%)                                                                                                                 
 & 3.69                                                                                                                  
 & 1.77                                                                                                                
 & 0.44                                                                                                                
 & -                                                                                                            
 &                                                                                                                        
 & 0.008                                                                                                                
 & 4.79                                                                                                          
 & 2.38                                                                                                               
 & -                                                                                                                 
\\                                                                                                                        
\hline\hline
                                                                                                              
\end{tabular}                                                                                                             
\begin{flushleft}                                                                                                         
\end{flushleft}                                                                                                           
\label{TableA}                                                                     
\end{table}

The bottom panel of Fig. \ref{fig3} shows the time evolution of the number of 3-motifs. In the original network this number is fluctuating and present a huge spike at day 79. In the case of the Bonferroni network, the time evolution is fluctuating less and no  spike is present. In summary our statistical validation procedure select a network characterized by properties that are much more stable than the original network. We hypothesize that the Bonferroni network is able to retain links whose social motivations are typically more pronounced than the ones left out from the original network, thus the Bonferroni network is a better proxy for the underlying social network than the original one. This is especially so for the strong ties characterizing the social networks for short time windows but increasingly for weaker ones too, as the time window is made longer. Our hypothesis is supported by the results we obtain for some important network metrics and for the census of the 3-motifs and their dynamics.

\begin{figure}
\begin{center}
\includegraphics[scale=0.6]{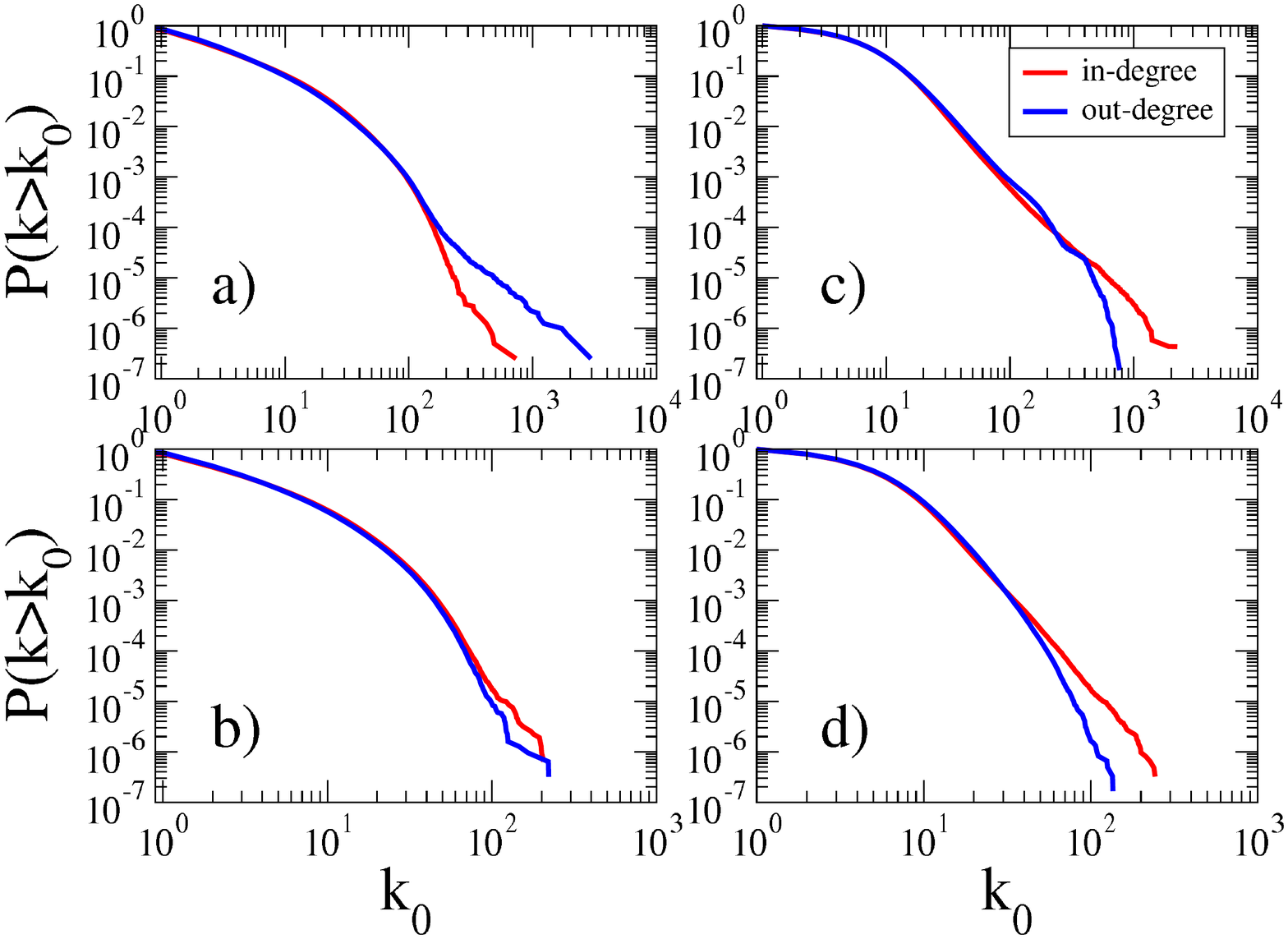}
\caption{Cumulative in-degree (red line) and out-degree (blue line) distributions. On left panels we show results for the Chinese dataset, entire period (109 days), subscribers only.  Panel a): original network. Panel b): Bonferroni network. On right panels we show results for the European dataset, entire period (212 days), subscribers only.  Panel c): original network. Panel d): Bonferroni network. A single occurrence with in-degree equals to 23348 is not shown in panel c) and a single occurrence with in-degree equals to 1131 is not shown in panel d).}  
\label{fig4} 
\end{center}
\end{figure}

In Fig. \ref{fig4} we show the cumulative in-degree and out-degree distributions for the Chinese datasets (subscribers only) for the entire period (109 days). The cumulative distributions are shown for the original network (panel a)) and for the Bonferroni network (panel b)). We observe a series of interesting differences in the cumulative distributions between the original and the Bonferroni networks. In the original network we observe subscribers with very large out-degree (of the order of 3,000) and in-degree (of the order 700). We also note that the tails of the in-degree and out-degree distributions are very pronounced and quite different the one from the other (with the out-degree distribution significantly more pronounced that the in-degree distribution). In the case of the Bonferroni network the in-degree and out-degree distributions are still showing pronounced tails but the largest degree is of the order of 200.  We also note that the tails of the in-degree and out-degree distributions are in the Bonferroni case similar, with the in-degree being only slightly more pronounced than the out-degree for very large degrees.

A similar pattern is observed also in the degree distributions of European data (see panels c) and d) of Fig. \ref{fig4}). However, we also note differences between the European and the Chinese distributions. Specifically, for the original network (panel c) of Fig. \ref{fig4}) the more pronounced tail is observed for the in-degree distribution in the European case whereas the opposite is observed in the Chinese case. We also note that in the European case the distributions of the Bonferroni networks show slightly different tails: The tail of the in-degree distribution is more pronounced than that of the out-degree. It is worth noting that similarly to what we observe for the Chinese dataset the maximal in-degree is close to 150 and the maximal out-degree is close to 250, if we do not consider the outlier characterized by a degree of 1131. The tails of the cumulative distributions of the in-degree and out-degree in the European case are well described by a power law decay with an exponent equals to 3.85 and 6.25 respectively \footnote{The exponent for out-degree distribution is pretty big. It decays so fast that it is difficult to distinguish between power law and exponential decay.}.

\section{3-motifs}

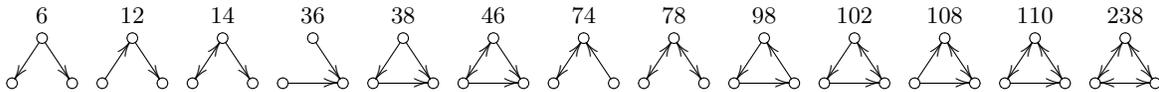
\begin{figure}[h]
  \centering
  \begin{xy}
    \POS (0,4) *{6}
    \POS (0,1) *\cir<2pt>{} ="a", (-4,-5)*\cir<2pt>{} ="b", (4,-5)*\cir<2pt>{} ="c"
    \POS "a" \ar @{->} "b"
    \POS "a" \ar @{->} "c"
    \POS (12,4) *{12}
    \POS (12,1) *\cir<2pt>{} ="a", (8,-5)*\cir<2pt>{} ="b", (16,-5)*\cir<2pt>{} ="c"
    \POS "a" \ar @{<-} "b"
    \POS "a" \ar @{->} "c"
    \POS (24,4) *{14}
    \POS (24,1) *\cir<2pt>{} ="a", (20,-5)*\cir<2pt>{} ="b", (28,-5)*\cir<2pt>{} ="c"
    \POS "a" \ar @{<->} "b"
    \POS "a" \ar @{->} "c"
    \POS (36,4) *{36}
    \POS (36,1) *\cir<2pt>{} ="a", (32,-5)*\cir<2pt>{} ="b", (40,-5)*\cir<2pt>{} ="c"
    \POS "a" \ar @{->} "c"
    \POS "b" \ar @{->} "c"
    \POS (48,4) *{38}
    \POS (48,1) *\cir<2pt>{} ="a", (44,-5)*\cir<2pt>{} ="b", (52,-5)*\cir<2pt>{} ="c"
    \POS "a" \ar @{->} "b"
    \POS "a" \ar @{->} "c"
    \POS "b" \ar @{->} "c"
    \POS (60,4) *{46}
    \POS (60,1) *\cir<2pt>{} ="a", (56,-5)*\cir<2pt>{} ="b", (64,-5)*\cir<2pt>{} ="c"
    \POS "a" \ar @{<->} "b"
    \POS "a" \ar @{->} "c"
    \POS "b" \ar @{->} "c"
    \POS (72,4) *{74}
    \POS (72,1) *\cir<2pt>{} ="a", (68,-5)*\cir<2pt>{} ="b", (76,-5)*\cir<2pt>{} ="c"
    \POS "a" \ar @{<->} "b"
    \POS "a" \ar @{<-} "c"
    \POS (84,4) *{78}
    \POS (84,1) *\cir<2pt>{} ="a", (80,-5)*\cir<2pt>{} ="b", (88,-5)*\cir<2pt>{} ="c"
    \POS "a" \ar @{<->} "b"
    \POS "a" \ar @{<->} "c"
    \POS (96,4) *{98}
    \POS (96,1) *\cir<2pt>{} ="a", (92,-5)*\cir<2pt>{} ="b", (100,-5)*\cir<2pt>{} ="c"
    \POS "a" \ar @{->} "b"
    \POS "a" \ar @{<-} "c"
    \POS "b" \ar @{->} "c"
    \POS (108,4) *{102}
    \POS (108,1) *\cir<2pt>{} ="a", (104,-5)*\cir<2pt>{} ="b", (112,-5)*\cir<2pt>{} ="c"
    \POS "a" \ar @{->} "b"
    \POS "a" \ar @{<->} "c"
    \POS "b" \ar @{->} "c"
    \POS (120,4) *{108}
    \POS (120,1) *\cir<2pt>{} ="a", (116,-5)*\cir<2pt>{} ="b", (124,-5)*\cir<2pt>{} ="c"
    \POS "a" \ar @{<-} "b"
    \POS "a" \ar @{<->} "c"
    \POS "b" \ar @{->} "c"
    \POS (132,4) *{110}
    \POS (132,1) *\cir<2pt>{} ="a", (128,-5)*\cir<2pt>{} ="b", (136,-5)*\cir<2pt>{} ="c"
    \POS "a" \ar @{<->} "b"
    \POS "a" \ar @{<->} "c"
    \POS "b" \ar @{->} "c"
    \POS (144,4) *{238}
    \POS (144,1) *\cir<2pt>{} ="a", (140,-5)*\cir<2pt>{} ="b", (148,-5)*\cir<2pt>{} ="c"
    \POS "a" \ar @{<->} "b"
    \POS "a" \ar @{<->} "c"
    \POS "b" \ar @{<->} "c"
  \end{xy}
  \caption{List of directed 3-motifs.} 
\label{Fig6}
\end{figure}

In Fig. \ref{Fig6} we show the 13 different types of 3-motifs that can be observed in a network. There are different ways to code the identity of these motifs. In the present paper we use the labeling of Milo et al. \cite{Alon2002}. We are interested in detecting the properties of communication 3-motifs in a social network using the mobile phone network of a specific mobile phone operator as a proxy for communications in the underlying social network.   

\begin{figure}
\begin{center}
\includegraphics[scale=0.8]{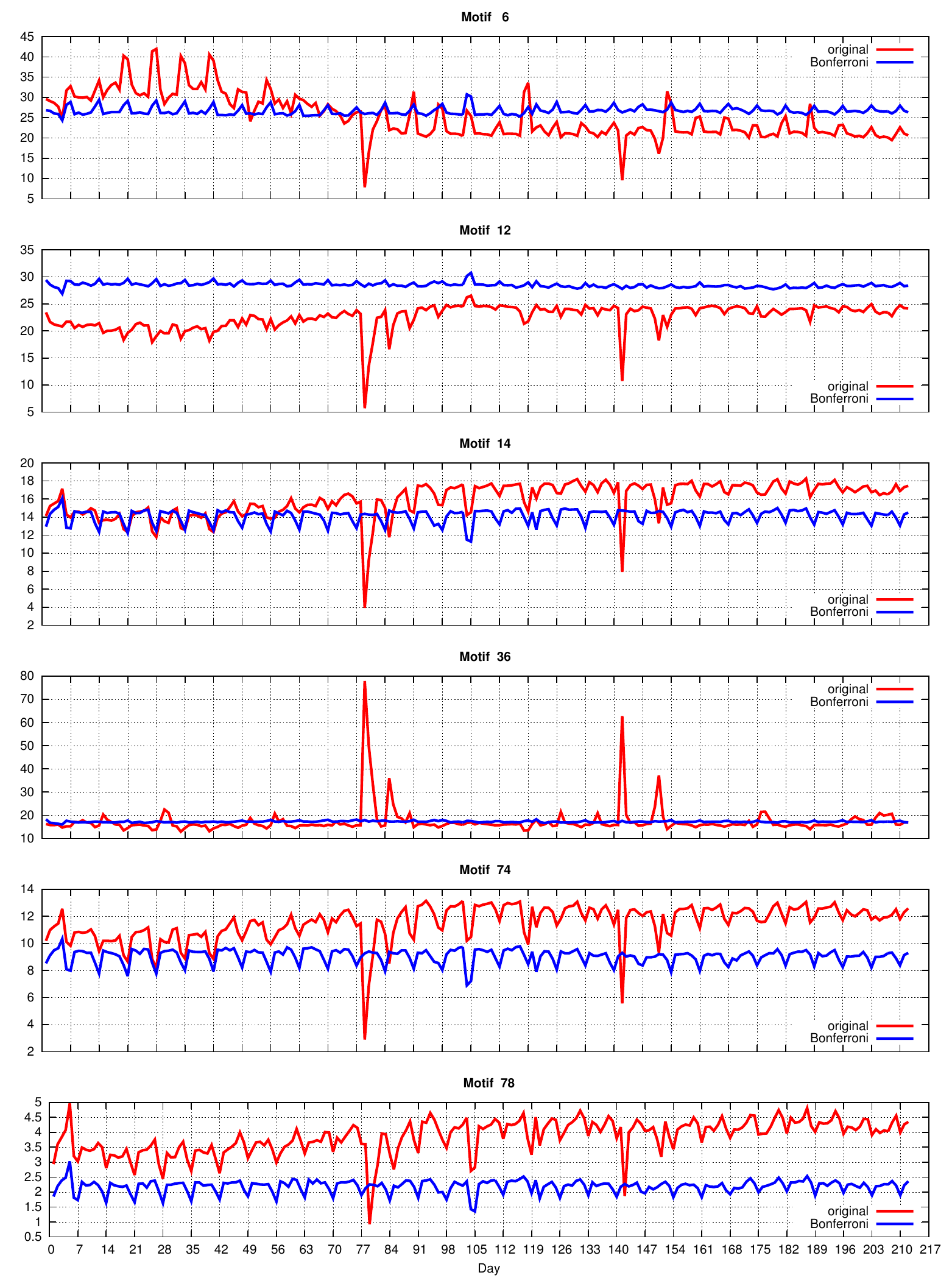}
\caption{Fraction of 3-motifs present in each day of the European data set. From top to bottom we have 3-motifs encoded as 6, 12, 14, 36, 74, and 78 respectively. The red line refers to the original networks whereas the blue line refers to the Bonferroni networks.}  
\label{fig7} 
\end{center}
\end{figure}

\begin{figure}
\begin{center}
\includegraphics[scale=0.8]{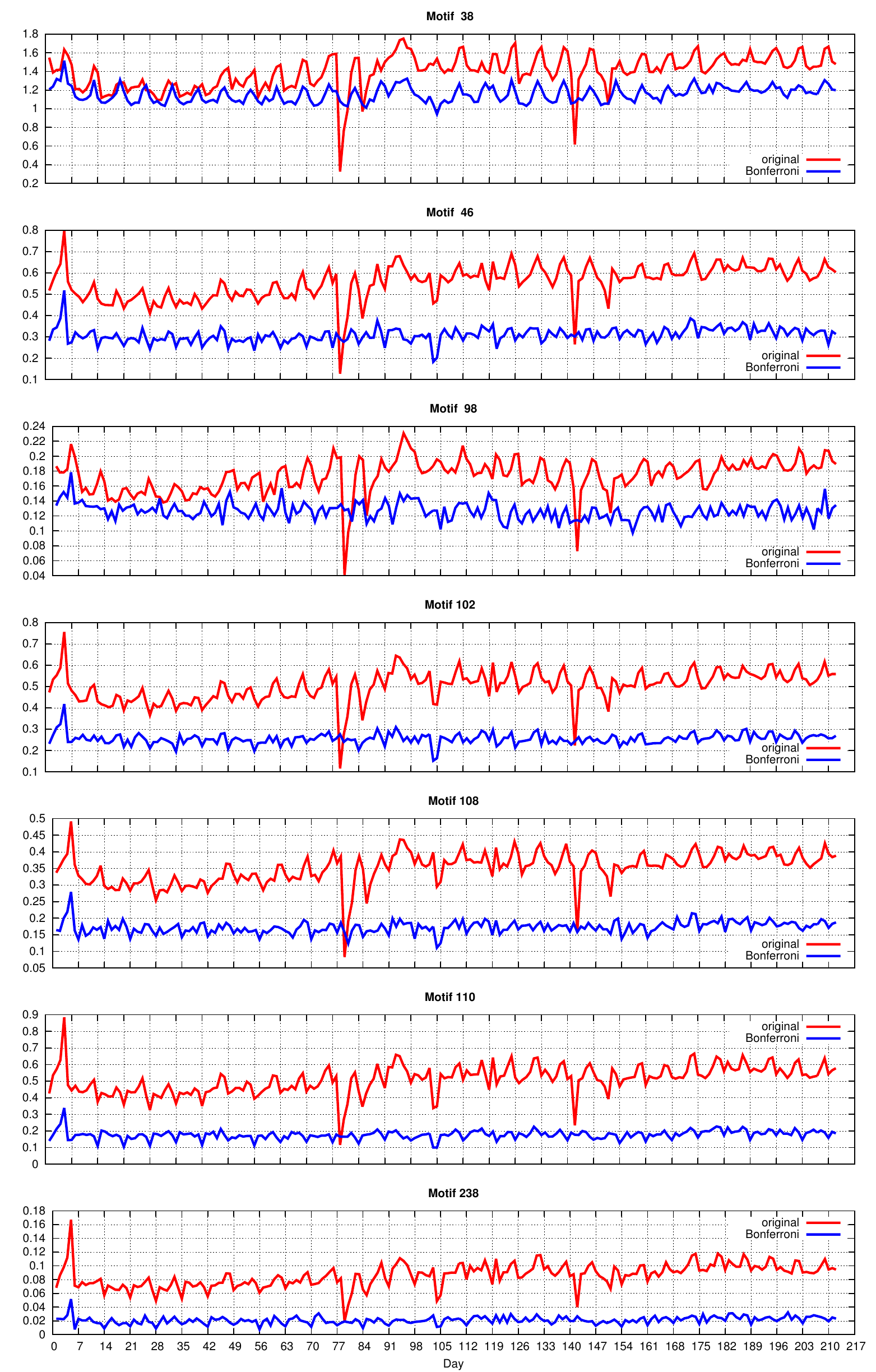}
\caption{Fraction of 3-motifs present in each day of the European data set. From top to bottom we have 3-motifs encoded as 38, 46, 98, 102, 108, 110, and 238 respectively. All these 3-motifs present a pair interaction for all pairs in one or both directions during the investigated day. The red line refers to the original networks whereas the blue line refers to the Bonferroni networks.}  
\label{fig8} 
\end{center}
\end{figure}

A direct inspection of Figures \ref{fig7} and \ref{fig8} shows that the estimation of the fraction of daily 3-motifs for the original network presents seasonalities of various frequencies and huge spikes localized at specific weeks. The seasonality is extremely pronounced for 3-motifs presenting only two of the three possible relationships (see the panels of Fig. \ref{fig7}). On the other hand the pattern observed in the Bonferroni networks is more stable and shows only a weekly seasonality and a small deviation occurring for some special days (day label 5 and 105 most probably related to big holidays periods). In the Bonferroni network, the weekly pattern is quite evident for the 3-motifs with two pair relationships (Fig. \ref{fig7}) whereas for triads with a triangle structure the weekly pattern is less evident especially in some case as, for example, for the 3-motif labeled as 98. 

We interpret these empirical results as supporting our hypothesis that the Bonferroni network is sampling relationships characterized by a strong social interaction whereas the original network also includes type of calls that are related to commercial or technical activities such as the ones typically performed by call centers. The presence of these activities can alter substantially counts of the triads because a node with a very large in-degree or out-degree participates in a large number of triads. This kind of spurious effect is clearly not observed in the Bonferroni network.

In Tables \ref{Table1} and \ref{Table2} we present a summary statistics of the fraction of 3-motifs observed in the daily, weekly and monthly original and Bonferroni networks for the Chinese and European datasets respectively. For each 3-motif and for each network we report the average value observed in real data $\mu$ and the average value $\mu_{\rm{rnd}}$ observed by randomly shuffling the network a large number of times while keeping fixed the in-degree and out-degree of each node and by maintaining the number of bidirectional relationships constant. The counting of the 3-motifs and the shuffling procedures were performed by using the FANMOD algorithm \cite{Fanmod}. In the Tables we also report the standard deviation observed in real data $\sigma$ and in shuffled data $\sigma_{\rm{rnd}}$.

For each 3-motif we evaluate a $z$-score defined as $z=(\mu-\mu_{\rm{rnd}})/\sigma$. This variable indicates the deviation of the observed average value from the average value obtained by random shuffling of the network in units of the standard deviation. We have decide to use this definition of $z$-score instead of the another possible one defined as $z_2=(\mu-\mu_{\rm{rnd}})/\sigma_{\rm{end}}$, because our definition is the most conservative one in the present case. In the Tables we highlight the average percent of a 3-motif in boldface character when its associated $z$-score is larger than 3 (a character $\bf{(+)}$ is following the average percent in this case) or smaller than -3 (a character $\bf{(-)}$ is following the average percent). Tables  \ref{Table1} and \ref{Table2} show that 3-motifs split in  two sets. The first one is the set of 3-motifs showing communication arcs only between two of the three pairs of nodes of the motif, i.e. 3-motifs encoded as 6, 12, 14, 36, 74, and 78. The second set is the set of 3-motifs with all pairs showing at least one communication link (3-motifs encoded as 38, 46, 98, 102, 108, 110, and 238). Tables  \ref{Table1} and \ref{Table2} show that for the first set of 3-motifs, the average percent of the 3-motifs is close to the value expected for random connections (6, 12 , and 36)  or less than expected for the 3-motifs 14, 74, and 78. On the other hand, for the second set of motifs (38, 46, 98, 102, 108, 110, and 238) all the 3-motifs are presenting an average percent which is higher than the value expected for random driven communications. In other words, the underlaying social structure and the communication style of the social actors over-express the 3-motifs characterized by triadic closure. This behavior is observed at daily, weekly and monthly time scale (with a pattern more pronounced when the time period used to build the network is longer) and it is observed both for the Chinese and the European datasets. 

The above cited results are qualitatively observed both for the original and the Bonferroni networks.  However, original and Bonferroni networks present values of the average percent of the 3-motifs which are quite different especially for weekly and monthly time periods. The difference is quite pronounced for 3-motifs of the first set (see, for example the average percent of 3-motif number 6 for the monthly networks). Our analysis of the time dependence of the average percent summarized in Figures \ref{fig7} and \ref{fig8} indicates that the results obtained for the Bonferroni networks are more robust and reliable than the results obtained for the original network and allows for a more detailed investigation of the process of formation and disappearance of these communication structures. In the next section we will analyze the process of formation of the communication 3-motifs in the most reliable setting which is the setting of the Bonferroni networks.

\section{Temporal evolution of communication 3-motifs}
Communication 3-motifs are continuously formed and disappearing over time. Here we primarily focus on the dynamics of the 3-motifs formation observed at a daily time scale. Specifically, we detect the Bonferroni network at day $k$ and at the two-days time interval beginning at day $k$ and we count and identify all the 3-motifs present in each network. The identification of each 3-motif is done by considering the identity of the three social actors composing it. In other words we keep memory of the fact that, for example, one 3-motif of type 6 is observed among subscribers with identity  $i$, $j$ and $k$. This is done to follow each 3-motif evolution during the increase of the monitoring time interval, which is primarily producing a network expansion \footnote{Indeed during the increase of the time interval used to obtain the Bonferroni network of a longer time period some links existing in the first Bonferroni network might also disappear due to absence of validation of the link in the second extended period of detection but not in the first period. The probability of disappearance of a link is pretty small but in a few cases such events occur.}. 

In Tables \ref{Table3} and \ref{Table4} we show the conditional probabilities for 3-motifs during a 1-day expansion of the time period used to determine the Bonferroni network. The starting network is computed for day $k$ whereas the target network is computed for a two days time interval including the previous one (days $k$ and $k+1$) for Chinese (Table \ref{Table3}) and European (Table \ref{Table4}) datasets. The networks refer to the cases of Bonferroni networks obtained from the records of subscribers. We highlight in boldface characters the entries with conditional probability higher than 0.05. By inspecting Tables \ref{Table3} and \ref{Table4} we note that the conditional probability $P(M_{\rm{II}}|M_{\rm{I}})$ shows the highest value in each row when $M_{\rm{II}}=M_{\rm{I}}$, i.e. when the 3-motif in the expanded network $M_{\rm{II}}$ is the same as the 3-motif in the starting one $M_{\rm{I}}$. This observation suggests that the detection of the 3-motif in Bonferroni network is pretty robust for all types of 3-motifs. The conditional probability is ranging between 0.395 (3-motif 98) to 0.916 (3-motif 238) and between 0.418 (3-motif 98) and 0.966 (3-motif 238) for the Chinese and European data respectively. It is worth noting that the less stable 3-motif is the one labeled as 98, which is a motif characterized by a directional flux of information among the 3 social actors. The second lowest value of the conditional probability $P(38|38)$ is observed for the other 3-motif which is a triad of unidirectional links. 

We also observe that the second largest value in each row of conditional probabilities is associated with a 3-motif pair requiring that a unidirectional link of the original 3-motif modifies into a bidirectional one in the arrival 3-motif. This observation suggests that the underlying communication process governing the 3-motif dynamics is primarily related to the probability of observing return calls (see $P(14|6)$, $P(14|12)$ and $P(74|12)$, $P(78|14)$, ..., etc) between two social actors.

\begin{figure}
\begin{center}
\includegraphics[scale=0.5]{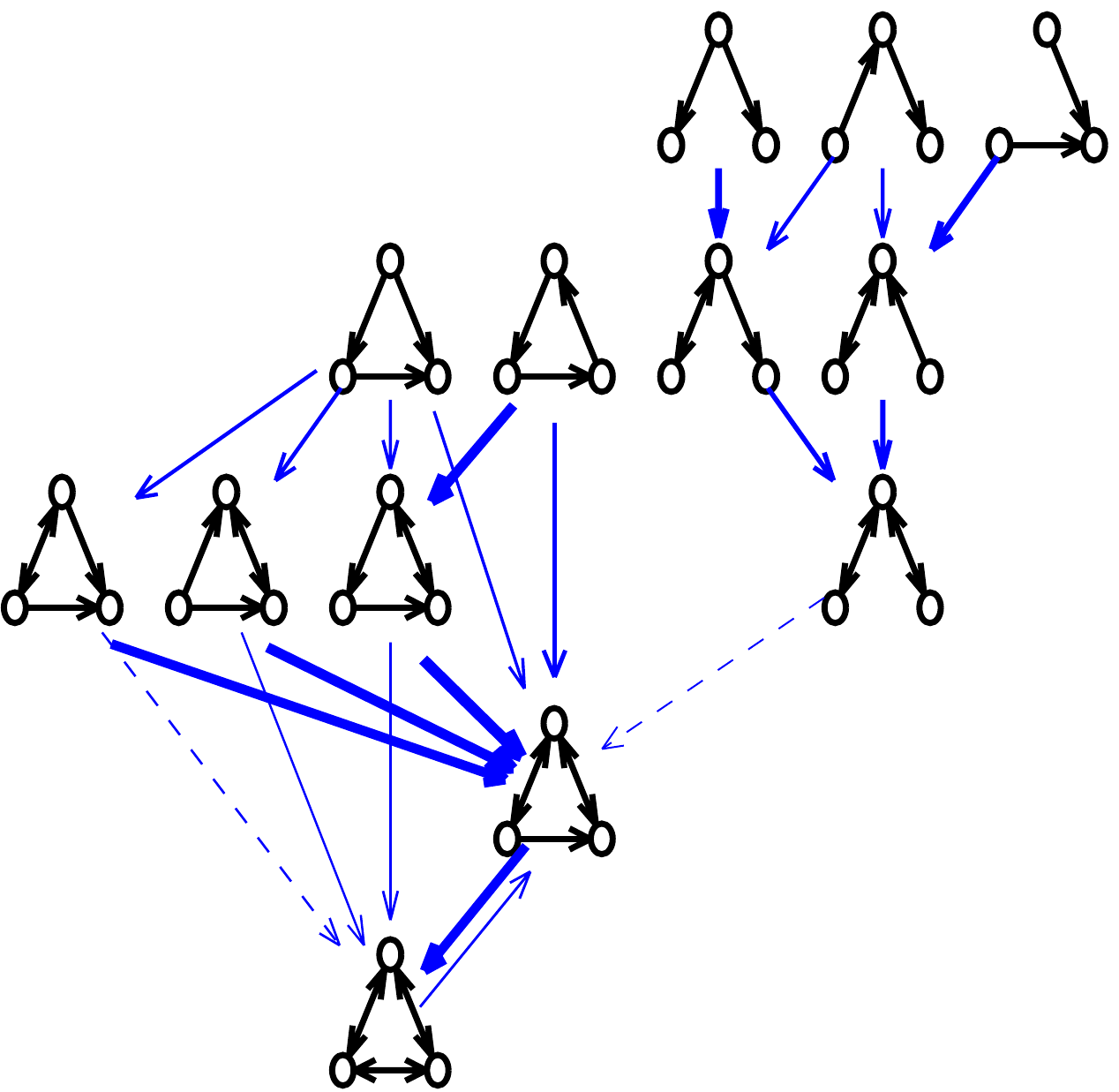}
\includegraphics[scale=0.5]{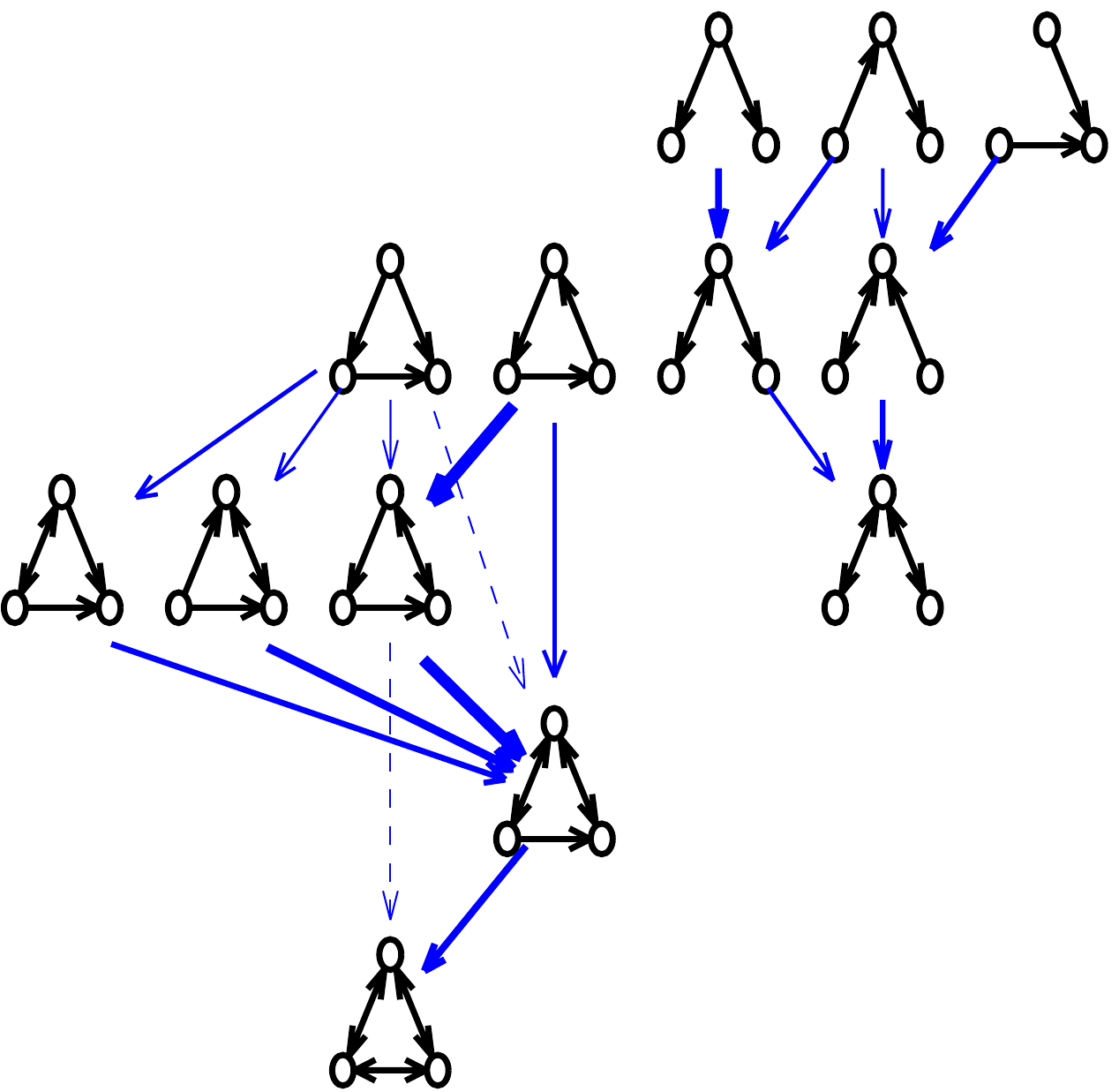}
\caption{Schematic representation of transition probabilities of 3-motifs when the network expand from a Monday network to a Monday-Tuesday network. Transition probabilities of more than 0.05 are indicated with blue arrows. The thickness of an arrow is proportional to the value of the conditional probability. The left and right panels refer to the Chinese and European datasets respectively.}  \label{fig9} 
\end{center}
\end{figure}

In Fig. \ref{fig9} we provide a schematic representation of the most relevant conditional probability among the different 3-motifs. In the panels of the figure we draw a line when the conditional probability from the originating to the arrival 3-motif exceed 5\%. For both the Bonferroni networks obtained from the 1-day expansion of the Chinese and European datasets we observe that the typical path of a 3-motif communication does not preferentially show triadic closure of open triangles but rather completion of reciprocal calls. In other words, the typical evolution path of a 3-motif, when the time interval used to detect the network extends from one day to two days, shows that just a small fraction  of 3-motifs evolve from open 3-motifs (i.e. 3-motifs with communication links detected only between two pairs of actors) to closed 3-motifs (i.e. 3-motifs with communication links detected between all three pairs of actors). Open 3-motifs preferentially tend to evolve towards bidirectional open 3-motifs, and only when links are fully reciprocated in the open 3-motif (motif 78) then the motif tends to evolve to a closed 3-motif---along the lines of triadic closure. 

\begin{figure}
\begin{center}
\begin{tabular}{cc}
\includegraphics[scale=0.3]{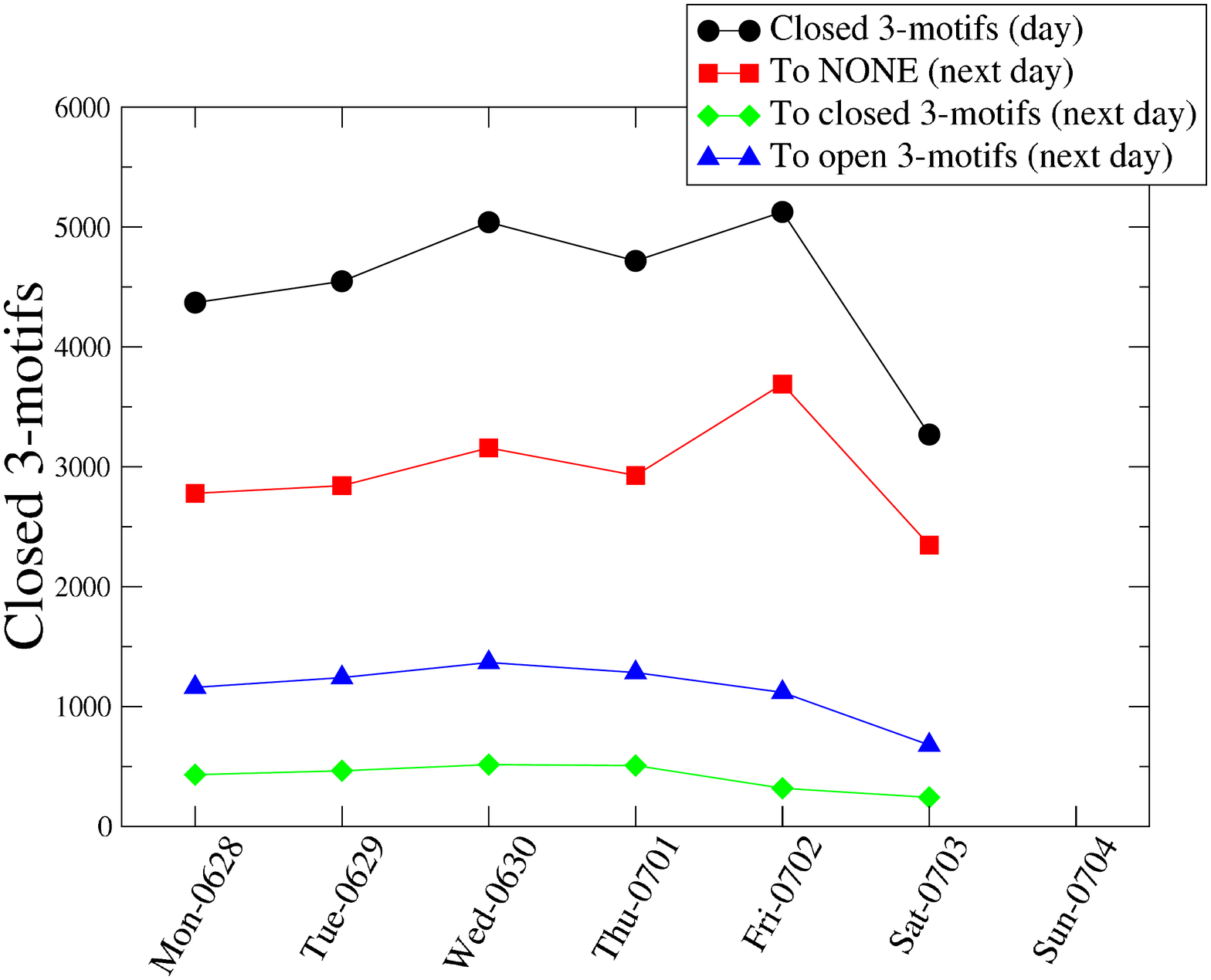} & \includegraphics[scale=0.3]{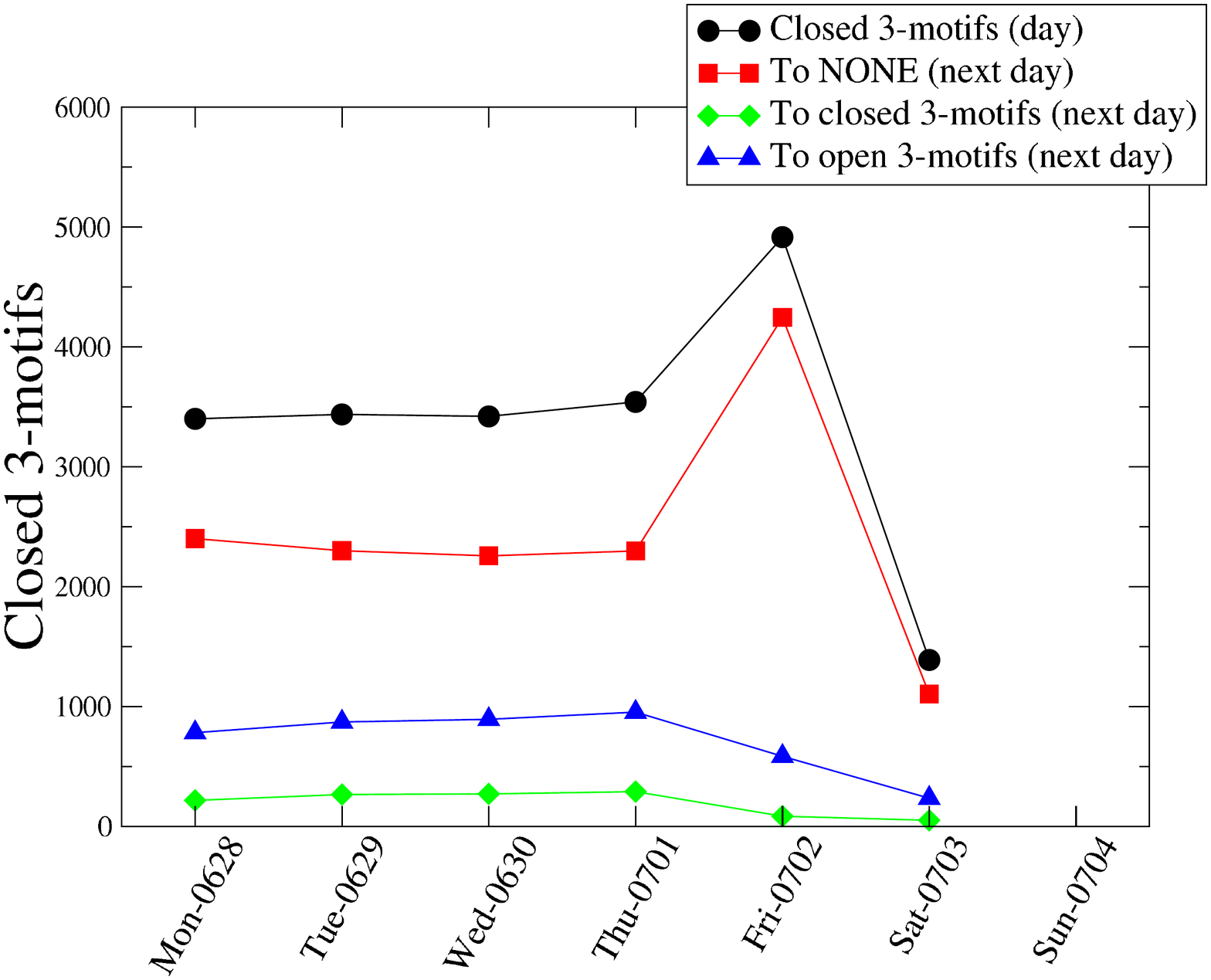}\\
\includegraphics[scale=0.3]{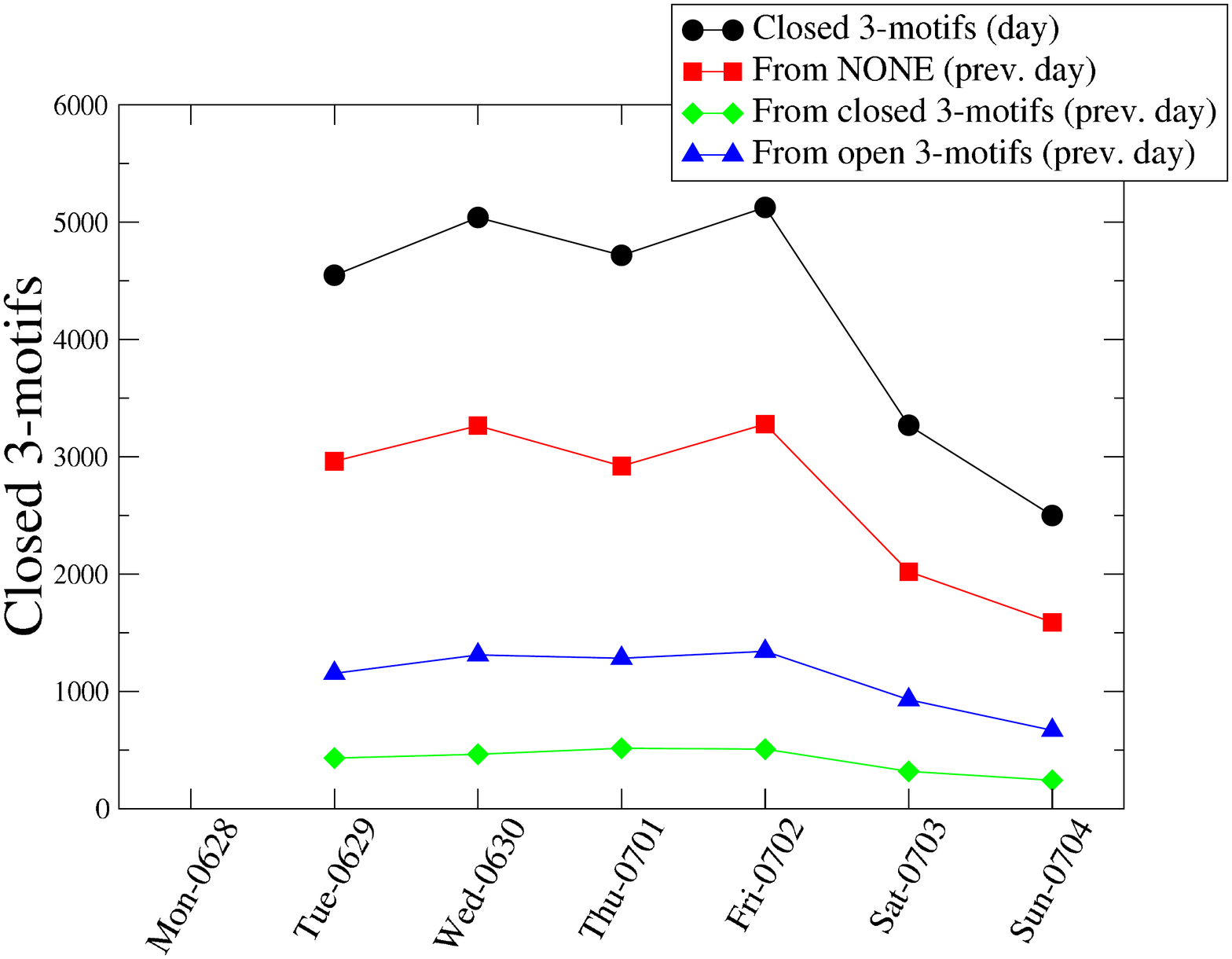} & \includegraphics[scale=0.3]{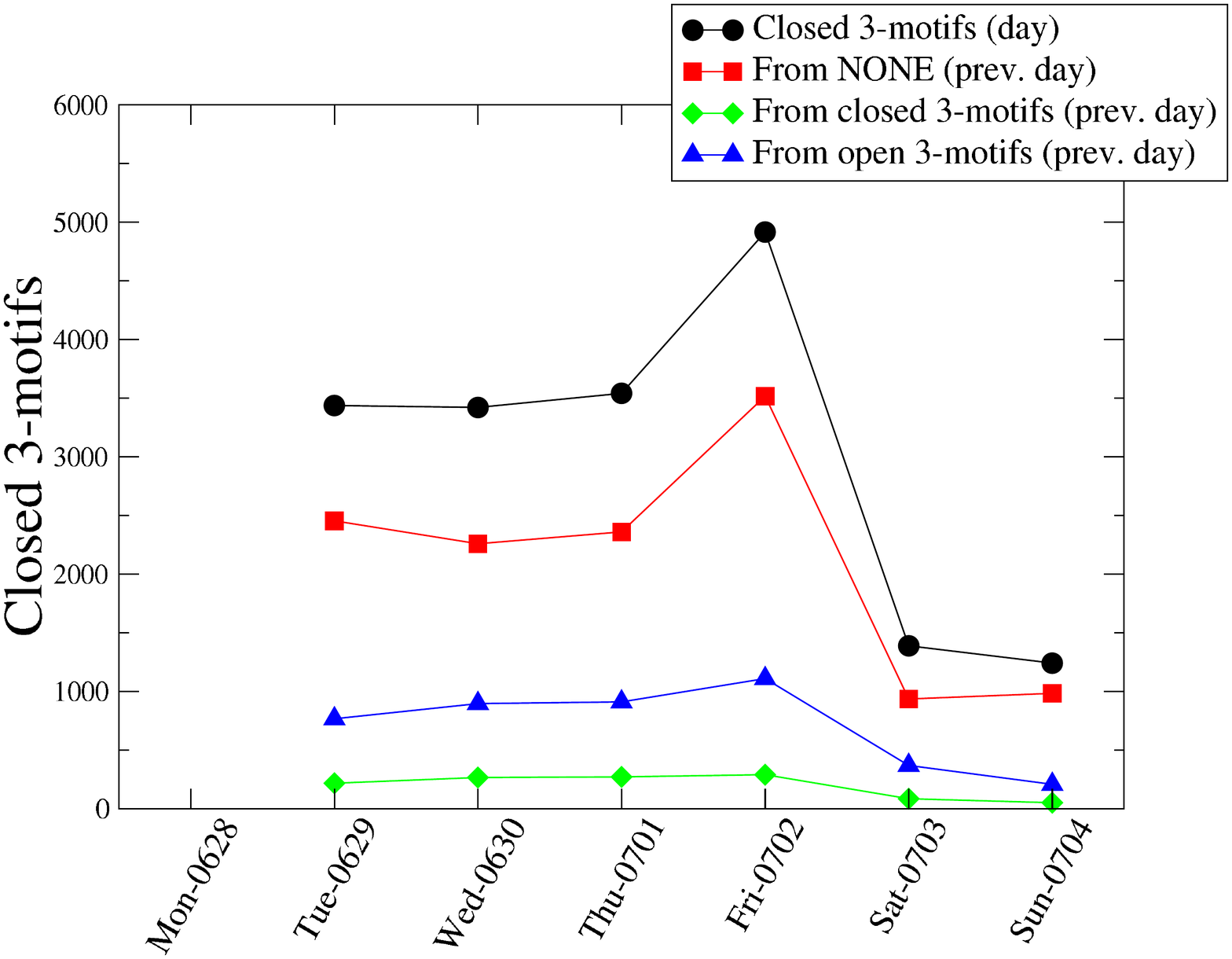}
\end{tabular}
\caption{Top panels: Evolution of closed 3-motifs for Chinese (left) and European (right) data across a week. Black circles indicate the total count of closed 3-motifs in the daily Bonferroni networks for the first six days of a week. Red rectangles indicate the part of these motifs that evolve, the day after, to node triplets that do not determine a 3-motif. Blue triangles (green diamonds) indicate the part of the total number of closed 3-motifs that evolve, the day after, to open 3-motifs (closed 3-motifs). Bottom panels: Formation of closed 3-motifs for Chinese (left) and European (right) data across a week. Black circles indicate the total count of closed 3-motifs in the daily Bonferroni networks for the last six days of a week. Red rectangles indicate the part of these motifs that emerge from node triplets that did not determine a 3-motif in the Bonferroni network of the day before. Blue triangles (green diamonds) indicate the part of the total number of closed 3-motifs that appeared as open 3-motifs (closed 3-motifs) in the Bonferroni network of the day before.}  
\label{fig10} 
\end{center}
\end{figure}

We interpret this observation as a manifestation of the fact that communication closed 3-motifs typically form at an intraday time scale. This interpretation is also supported by the results reported in Fig.\ref{fig10}, where we look at the detail of formation and evolution of closed 3-motifs from one day to the next one, without varying the time window and without distinguishing between the different closed 3-motifs and the different open 3-motifs. On average, more than $2/3$ of the closed three motifs observed in the Bonferroni network of a given day come from unconnected triples of nodes in the Bonferroni network of the previous day, and evolve to unconnected triplets of nodes in the Bonferroni network of the following day (red rectangles in the figure). On the other hand, the number of closed 3-motifs that form from (evolve to) open 3-motifs in the Bonferroni network of the previous (following) day is about $1/4$ of the total (blue triangles in the figure). Such an erratic pattern suggests that a large fraction of closed 3-motifs that appear in a daily Bonferroni network occur due to contingent reasons of communication that develop at an intra-day time scale, e.g., the peak of closed 3-motifs observed on Friday may be due to the need of people to coordinate their social activities.

%*************

\section{Conclusions}

In this paper we have adapted and applied a filtering procedure to a directed communication network. This filtering procedure %The filtering procedure 
is based on a statistical validation performed by using multiple hypothesis test correction. We hypothesize that the links detected in the directed Bonferroni communication networks describe the 
relevant ties of the underlying social structure originating the communication. 
We test our hypothesis by comparing basic statistics of the original and the Bonferroni networks and conclude that the latter one is much more realistic as it removes spurious links related to non-social interactions. Furthermore, we investigate the relative frequency of 3-motifs in two large sets of mobile communication data recorded in two different countries of two distinct continents. In both cases we verify that the frequency profile of the 3-motifs of the directed Bonferroni communication networks are much more stable over time than the frequency profile of the original network. We believe that this empirical observation supports the hypothesis of Bonferroni networks being good proxies of strong ties of social origin.  

After having verified the robustness and reliability of our statistical filtering procedure, we have investigated the time evolution of the communication 3-motifs. Our results show that communication 3-motifs characterized by triadic closure form frequently at an intraday time scale. 
On the other hand, open 3-motifs (i.e. 3-motifs with links detected only between two pairs of subscribers) primarily evolve to other open 3-motifs with a higher number of reciprocated calls. In fact, the preferential path of evolution of open 3-motifs shows that the open 3-motif evolves to a closed triad with a sizeable conditional probability when all the calls of the open 3-motif are reciprocated. 

We interpret these results as an evidence for 
the fact that correctly sampled mobile call records are reflecting rapid communication interactions of an underlying social structure that forms and dissolves over a longer time scale. In other words, the time scales of the communication network and of the social network are quite distinct with the first lasting usually less than a day and the second requiring months or years. Under this interpretation we conclude that the triadic closure process is governed by distinct rules in communication and in social networks.      

Acknowledgments - This work was partially supported by the National Natural Science Foundation of China (11205057), the Humanities and Social Sciences Fund of the Ministry of Education of China (09YJCZH040), the Ph.D. Programs Foundation of Ministry of Education of China (20120074120028), the Fok Ying Tong Education Foundation (132013), and the Fundamental Research Funds for the Central Universities.

\begin{table}[ht]                                                                                                         
\caption{Statistics of 3-motifs for the Chinese data. Subscribers only.  The networks investigated are the original network and the Bonferroni network. Average value observed in real data $\mu$ and the average value $\mu_{\rm{rnd}}$ observed by randomly shuffling the network.  We also report the standard deviation observed in real data $\sigma$ and in shuffled data $\sigma_{\rm{rnd}}$. Values are given in percent. Daily, weekly and monthly time periods. Values labeled in boldface indicate positive (+) or negative (-) $z$-score values larger than 3 in absolute value. The $z$-score is computed as $z=(\mu-\mu_{\rm{rnd}})/\sigma$.}                                                       
\centering                                                                                                                
\begin{tabular}{*{11}{c}}                                                                                                 
\hline\hline                                                                                                              
 & & \multicolumn{4}{c}{Original}                                                                                         
 & & \multicolumn{4}{c}{Bonferroni}\\                                                                                     
\cline{3-6}
\cline{8-11}                                                                                                   
 & & $\mu$ & $\sigma$ & $\mu_{\rm{\rm{rnd}}}$ & $\sigma_{\rm{\rm{rnd}}}$                                                            
 & & $\mu$ & $\sigma$ & $\mu_{\rm{\rm{rnd}}}$ & $\sigma_{\rm{\rm{rnd}}}$                                                            
\\
\hline\hline                                                                                                              
\end{tabular}                                                                                                             
\begin{flushleft}                                                                                                         
%\small $\mu$ is the average percentage of the number of one motif. $\sigma$ is the standard deviation of the percentages.
\end{flushleft}                                                                                                           
\label{Table1}                                                                     
\end{table}

\begin{table}[ht]                                                                                                         
\caption{Statistics of 3-motifs European data. Subscribers only.  The networks investigated are the original network and the Bonferroni network. Average value observed in real data $\mu$ and the average value $\mu_{\rm{rnd}}$ observed by randomly shuffling the network.  We also report the standard deviation observed in real data $\sigma$ and in shuffled data $\sigma_{\rm{rnd}}$. Values are given in percent. Daily, weekly and monthly time periods. Values labeled in boldface indicate positive (+) or negative (-) $z$-score values larger than 3 in absolute value. The $z$-score is computed as $z=(\mu-\mu_{\rm{rnd}})/\sigma$.}                                                         
\centering                                                                                                                
\begin{tabular}{*{11}{c}}                                                                                                 
\hline\hline                                                                                                              
 & & \multicolumn{4}{c}{Original}                                                                                         
 & & \multicolumn{4}{c}{Bonferroni}\\                                                                                     
\cline{3-6}\cline{8-11}                                                                                                   
 & & $\mu$ & $\sigma$ & $\mu_{\rm{\rm{rnd}}}$ & $\sigma_{\rm{\rm{rnd}}}$                                                            
 & & $\mu$ & $\sigma$ & $\mu_{\rm{\rm{rnd}}}$ & $\sigma_{\rm{\rm{rnd}}}$                                                            
\\
\hline\hline                                                                                                              
\end{tabular}                                                                                                             
\begin{flushleft}                                                                                                         
%\small $\mu$ is the average percentage of the number of one motif. $\sigma$ is the standard deviation of the percentages.
\end{flushleft}                                                                                                           
\label{Table2}                                                                       
\end{table}

\begin{table}[ht]                                                                         
\caption{Conditional probabilities for 3-motifs during network 1-day expansion. The starting network is computed for day k $\rightarrow$ whereas the target network is computed for a two days time interval including the previous one (days k and k+1) for Chinese data. Subscribers only, Bonferroni networks.}
\centering                                                                                
\begin{tabular}{*{15}{c}}                                                                 
\hline\hline    
& 6 & 12 & 14 & 36 & 38 & 46 & 74 & 78 & 98 & 102 & 108 & 110 & 238 & ~ \\
\hline\hline                                                                              
\end{tabular}                                                                             
\label{Table3}                                        
\end{table}

\begin{table}[ht]                                                                       
\caption{Conditional probabilities for 3-motifs during network 1-day expansion. The starting network is computed for day k $\rightarrow$ whereas the target network is computed for a two days time interval including the previous one (days k and k+1) for European data. Subscribers only, Bonferroni networks.}
\centering                                                                              
\begin{tabular}{*{15}{c}}                                                               
\hline\hline                                                                            
& 6 & 12 & 14 & 36 & 38 & 46 & 74 & 78 & 98 & 102 & 108 & 110 & 238 & ~ \\
\hline\hline                                                                            
\end{tabular}                                                                           
\label{Table4}                                        
\end{table}    

\end{document}